\documentclass[aps,prd,amsmath,amssymb,preprintnumbers,nofootinbib,a4paper,preprint,floatfix]{revtex4-1}
\usepackage{amsthm}
\usepackage{amssymb}
\usepackage{graphicx}
\usepackage{color}
\newcommand{\beq}{\begin{eqnarray}}
\newcommand{\eeq}{\end{eqnarray}}

\newcommand{\bmp}{\noindent\begin{minipage}{16cm}}
\newcommand{\emp}{\end{minipage}\vskip 7mm} % 7mm untightened

\usepackage{dcolumn}% Align table columns on decimal point
\usepackage{bm}% bold math
\usepackage{bbm}
\usepackage{subfigure}
\usepackage{pxfonts}
\usepackage{youngtab}
\usepackage{slashed}
\usepackage{graphicx}
\usepackage{multirow}
\usepackage{epsfig}

\usepackage[%dvips,
bookmarks]{hyperref}

\def\wick#1{\setbox2=\hbox{$\displaystyle#1$}
    \setbox3=\null\ht3=3.0pt\dp3=0.0pt\wd3=20.0pt
    #1\kern-\wd2\kern3.0pt\raise11.0pt\vbox{\hrule height0.3pt
    \hbox{\vrule width0.3pt\box3\vrule width0.3pt}}\kern-24.0pt\kern\wd2}

\def\longwick#1{\setbox2=\hbox{$\displaystyle#1$}
    \setbox3=\null\ht3=3.0pt\dp3=0.0pt\wd3=27.0pt
    #1\kern-\wd2\kern3.0pt\raise11.0pt\vbox{\hrule height0.3pt
    \hbox{\vrule width0.3pt\box3\vrule width0.3pt}}\kern-31.0pt\kern\wd2}

\def\verylongwick#1{\setbox2=\hbox{$\displaystyle#1$}
    \setbox3=\null\ht3=3.0pt\dp3=0.0pt\wd3=43.0pt
    #1\kern-\wd2\kern3.0pt\raise11.0pt\vbox{\hrule height0.3pt
    \hbox{\vrule width0.3pt\box3\vrule width0.3pt}}\kern-47.0pt\kern\wd2}

\definecolor{bluc}{cmyk}{1,1,0,0.1}
\definecolor{rossoCP3}{cmyk}{0,.88,.77,.40}
\definecolor{rosso}{cmyk}{0,1,1,0.4}
\definecolor{rossos}{cmyk}{0,1,1,0.55}
\definecolor{rossoc}{cmyk}{0,1,1,0.2}
\definecolor{verdes}{cmyk}{0.92,0,0.59,0.4}

\hypersetup{colorlinks, bookmarksopen, bookmarksnumbered,
citecolor=verdes, linkcolor=bluc, pdfstartview=FitH, urlcolor=rossos}

\theoremstyle{definition}

\theoremstyle{plain}

\def\lsim{\mathrel{\rlap{\lower4pt\hbox{\hskip1pt$\sim$}}
    \raise1pt\hbox{$<$}}}                % less than or approx. symbol
\def\gsim{\mathrel{\rlap{\lower4pt\hbox{\hskip1pt$\sim$}}
    \raise1pt\hbox{$>$}}}                % greater than or approx. symbol

\newcommand{\ba}{\begin{eqnarray}}
\newcommand{\ea}{\end{eqnarray}}
\newcommand{\be}{\begin{equation}}
\newcommand{\ee}{\end{equation}}
\newcommand{\bd}{\begin{displaymath}}
\newcommand{\ed}{\end{displaymath}}
\newcommand{\een}{\nonumber\end{equation}}
\newcommand{\bea}{\begin{eqnarray}}
\newcommand{\eean}{\nonumber\end{eqnarray}}
\newcommand{\eea}{\end{eqnarray}}

\def\cite#1{\citep{#1}}

\newcommand{\old}[1]{}

\begin{document}

\title{\Large  \color{rossoCP3} Approximating the Ising model  on
  fractal lattices \\ of dimension below two}
\author{Alessandro Codello$^{\color{rossoCP3}{\varheartsuit}}$}\email{codello@cp3-origins.net}
\author{Vincent Drach$^{\color{rossoCP3}{\varheartsuit}}$}\email{drach@cp3-origins.net} 
\author{Ari Hietanen$^{\color{rossoCP3}{\varheartsuit}}$}\email{hietanen@cp3-origins.net}

\affiliation{
\vspace{5mm}
{$^{\color{rossoCP3}{\varheartsuit}}${ \color{rossoCP3}  \rm  CP}$^{\color{rossoCP3} \bf 3}${\color{rossoCP3}\rm-Origins} \& the
  {\color{rossoCP3} \rm Danish IAS}, University of Southern Denmark, Campusvej 55, DK-5230 Odense M, Denmark}
}
 %%%%%%%%%%%%%%%%%%%%%%%%%%%%%%%%%%%%%%%%%%%%%%%%%%%%%%%%%%%%%%%%%%%%%%%%%%%%%%%%%%%%%%

\begin{abstract}

We construct periodic approximations to the free energies of Ising models
on fractal lattices of dimension smaller than two, in the case of zero
external magnetic field,
%. The result is obtained, as the limit of the exact free energies of the Ising model on periodic approximations,
using a generalization of the combinatorial method of Feynman and Vodvickenko.
%
%Internal energy, specific heat and entropy can be exactly evaluated at each step of the iteration. % any of the lattice successively approximate the fractal.   
%
Our procedure is applicable to any fractal obtained by the removal of sites of a periodic two dimensional lattice. %, like Archimedean or Laves lattices.
As a first application, we compute estimates for the critical
temperatures of many different Sierpinski carpets and we compare them to
known Monte Carlo estimates.
The results show that our method is capable of determining the
critical temperature with, possibly, arbitrary accuracy and paves the
way to determine $T_c$ for any fractal of dimension below two.
Critical exponents are more difficult to determine since the free energy of any periodic approximation still has a logarithmic singularity at the critical point implying $\alpha = 0$.  We also compute the correlation length as a function of the temperature and extract the relative critical exponent.
We find $\nu=1$ for all periodic approximation, as expected from universality.
%and they cast doubts on the ability for any method based on periodic approximations to determine the fractal critical exponents.
%
%Finally, we discuss the implications that our result offers regarding the problem of estimating the lower critical dimension for the Ising universality class.
\\
\\
{\textit{Preprint}} : CP3-Origins-2015-014 DNRF90 \& DIAS-2015-14

\end{abstract}
\maketitle

%\tableofcontents

%%%%%%%%%%%%%
\section{Introduction}
%%%%%%%%%%%%%

\paragraph*{Ising on fractals.}
The well known exact solutions of the Ising model in one and two
dimensions are the only exact solutions we have to
date~\cite{Ising:1925em,Onsager_1944}. Ising models on fractals of dimension between one and two are
natural possible candidates to enter this restricted group of solvable
models. Actually, we already have solutions on some fractals, those
with finite ramification number like the Sierpinski gasket, but these
are of limited interest since they resemble the one dimensional case
as they do not possess any phase transition at finite temperature \cite{gefen_1984}.
Instead, a fractal with infinite ramification number as the
Sierpinski carpet, which we know has a non-zero critical temperature \cite{Shinoda_2002,vezzani:2003aa}, has to date been studied mostly numerically, and few analytical studies are available. In this paper we try to fill this gap presenting an analytical study of the Ising model on fractals of dimension below two, which include both the gasket and the carpet. We present a method in principle able to determine the critical temperatures exactly for all these fractals. Our approach is based on approximating the Ising model on non-periodic fractal lattices with a sequence of Ising models on periodic lattices. We do this by exploiting our ability to readily solve the two dimensional Ising model on an arbitrary periodic lattice using an extension of the combinatorial method of Feynman--Vodvickenko \cite{Vodvicenko_1965,Feynman_1972,Codello2010}.

\paragraph*{Universality.}

The understanding of universality classes in dimension equal or above
two is now quite robust, in particular for system with
$\mathbb{Z}_2$ symmetry \cite{Codello:2012sc,El-Showk:2013nia}. Less
clear is the situation in dimension below two and greater than
one. Continuous
methods~\cite{Guida:1998bx,Ballhausen:2003gx}
give a fairly good description, and in some dimensions real space
renormalization group studies are available~\cite{Bonnier_1988,Monceau_2003}, but an
explicit solution of the Ising
model in some fractal case will provide strong indication regarding the
reliability or not of continuous methods in dimension below two.
In fact it is not completely clear if there is a difference between
the values of critical exponents one can obtain with continuous
methods, which usually make a continuation of the integer number of
dimensions to {\it fractional} values, and the actual values obtained
studying the analogous systems defined directly on lattices of
non-integer {\it fractal} dimension.
Another open question is if there is a lower critical
dimension for the $\mathbb{Z}_2$ universality class, or even if this concept is
well defined since it might be that  universality depends on the fine
details of the fractal \cite{gefen:1980aa}.  To clarify all these question, a better
understanding of the Ising model in fractal dimension, which is a
representative of the $\mathbb{Z}_2$ universality class, will be of the utmost significance. 
%or if this is one (obviously we are speaking of fractals with
%infinite ramification number).
A further reason why the Ising model universality class is interesting
in dimension below two is because it is the only non-trivial one in the
family of the $O(N)$ models due to the generalised Mermin-Wagner theorem~\cite{Cassi_1992}.

\paragraph*{Summary of the paper.}
In Section II we explain how to generalize the combinatorial method of Feynman and Vodvickenko
to arbitrary periodic lattices.
Then, in Section III, we explain how to apply it 
to approximate fractals. After some analytical results, we turn to
numerical methods to extract the approximate critical temperatures and correlation lengths for many
different fractals. We
finally compare our results with the numerous existing Monte Carlo
estimates available in the literature and we briefly discuss possible
future applications of our method. 

%previous version :
%After reviewing the combinatorial method of Feynman and
%odvickenko with its extension to arbitrary periodic lattices in Section I, we focus on Sierpinski carpet(s), i.e. all fracta%l constructed by eliminating sites in a L × L square lattice, in Section II, which we solve exactly at any finite iteration %k. Then we propose some ana- lytical insight, and then, due to the increasing analytical complexity, we turn to numerical me%thods to extract the exact critical temperatures for a large number of finite iteration fractals, which we use to estimate t%he true limiting temperatures of the full fractals. We then compare our results with the numerous existing Monte Carlo result% available in the literature and we briefly discuss possible future applications of our method.

%%%%%%%%%%%%%%%%%%%%%%
\section{Solution on arbitrary periodic lattices}
%%%%%%%%%%%%%%%%%%%%%%

%%%%%%%%%%%%%%%%%%%%%
\subsection{The model}
%%%%%%%%%%%%%%%%%%%%%

\paragraph*{Definitions.}
We briefly review the definitions that specify the model.
% and the basic steps involved in the combinatorial solution of it.
We consider an arbitrary periodic lattice $\Lambda$ where at every lattice site $i$ there is a spin variable
$\sigma_{i}\in\{-1,1\}$ and we define a microstate by a spin
configuration $\{\sigma\}$. We assume nearest neighbour interactions
so that the energy of a given spin configuration is given by 
\begin{equation}
E\{\sigma\}=-J\sum_{\left\langle i,j\right\rangle }\sigma_{i}\sigma_{j}\,.\label{defising}
\end{equation}
If $J>0$ the interaction is ferromagnetic, while it is antiferromagnetic
if $J<0$. The partition function is the sum over all spin configurations weighted by the Boltzmann-Gibbs
factor
\begin{equation}
Z_{\Lambda}=\sum_{\{\sigma\}}\exp \Big\{\beta \sum_{\left\langle i,j\right\rangle }\sigma_{i}\sigma_{j}\Big\}\,.\label{part}
\end{equation}
We define $\beta=1/k_{B}T$, where $k_{B}$ is the Boltzmann constant, and we set $J=1$ since we are going to consider only ferromagnetic interactions.

\paragraph*{High temperature expansion.}

Using the high temperature expansion, the partition function (\ref{part})
can be rewritten as
\begin{equation}
Z_{\Lambda}=2^{N}\left(\cosh \beta \right)^{N_l}\Phi_{\Lambda}(v)\,,
\label{2.1}
\end{equation}
where $v=\tanh \beta$ and $N\equiv N_s$ is the total number of lattice sites while $N_l$ is the total number of links.
%per lattice site.
The function $\Phi_{\Lambda}(v)$ is the generating function of the numbers
% $G_{n}$ (we define $G_{0}=1$),
which count the graphs with even vertices of a given length
% $n$ 
that can be drawn on the lattice $\Lambda$. In this way, the problem of
solving the Ising model on $\Lambda$ is reduced to the
combinatorial problem of counting even closed graphs on $\Lambda$.
In the thermodynamic limit it is the function $\Phi_\Lambda(v)$ which develops
the non-analyticity that characterises the continuous phase transition.
For this reason in the following we will focus on it and disregard the pre--factors appearing in (\ref{2.1}).

For high temperature expansion studies of the Ising model on fractals,
see~\cite{Fabio_1994}. As explained in the next section, we instead resum the
high temperature series by generalizing the approach of Feynman--Vodvickenko.

%%%%%%%%%%%%%%%%%%%%%
\subsection{Feynman--Vodvickenko method}\label{sub:method}
%%%%%%%%%%%%%%%%%%%%%

\paragraph*{Exact solution on arbitrary periodic lattices.}
%
%If we do not care about the fact that we have to count only over closed
%graphs with even vertices, we can use the standard combinatorial property
%that the sum over all (possibly disconnected) closed graphs, is given
%by the exponential of the sum over only connected closed graphs. But
%connected closed graphs can be seen as closed random walk paths. If
%we find a way to consider in the sum only closed paths with even vertices,
%we have a way to calculate the generating function $\Phi_{\Lambda}(v)$
%simply by counting closed random walks paths. 
Feynman \cite{Feynman_1972} and Vodvicenko \cite{Vodvicenko_1965}
introduced a trick to reduce the problem of counting closed graphs to a random walk
problem. More precisely, the generating function $\Phi_{\Lambda}(v)$
can be computed by counting closed weighted random walks paths on
$\Lambda$, where the weights are complex amplitudes constructed so
that the mapping from the high temperature expansion
to the random walk problem works out correctly \cite{Kac_Ward_1952,Sherman_1960,Morita_1985,Codello2010}.

From the knowledge of the transition matrix $\mathbb{W}_{\Lambda}$ of the random walk problem,
one can determine the explicit form of the generating function from the following relation \cite{Codello2010}:
\begin{equation}
\Phi_{\Lambda}(v)=\exp\left\{ \frac{N}{2}\int\frac{d^{2}k}{(2\pi)^{2}}\log\,\textrm{det}\left[\mathbb{I}-v\,\mathbb{W}_{\Lambda}(\mathbf{k})\right]\right\} \,,
\end{equation}
where the $\mathbf{k} = (k_{x},k_{y})$ integration is over the region $0\leq k_{x}\leq2\pi$ and $0\leq k_{y}\leq2\pi$.
The singular non-trivial part of the free energy for spin $\beta f_{\Lambda}= -\frac{1}{N}\log \Phi_{\Lambda}$
for the lattice $\Lambda$, can finally be written as
%
%\begin{widetext}
\begin{equation}
\beta f_{\Lambda}(v) = 
%-p\log2-\frac{z}{2}\log(1+v^{2})
-\frac{1}{2}\int\frac{d^{2}k}{(2\pi)^{2}}\,\textrm{log}\,P_\Lambda(v,\mathbf{k})\,,
\label{logdet}
\end{equation}
%\end{widetext}
%
where we have defined the determinant
\begin{equation}
P_\Lambda(v,\mathbf{k}) = \textrm{det}\left[\mathbb{I}-v\,\mathbb{W}_\Lambda(\mathbf{k})\right]\,.
\label{P}
\end{equation}
The matrices $\mathbb{W}_\Lambda(\mathbf{k})$ are $m\times m$ matrices with $m=s\times l$,
where $s$ is the number of sites in  the basic tile and $l$ is the total number of links in the basic tile.
%Here $z$ is the coordination number of the lattice $\Lambda$, while $p$ is the number of lattice sites in the unit cell of $\Lambda$.\\

\paragraph*{Critical temperature.}
If a phase transition takes place, the critical temperature can be determined as the real solution of
\begin{equation}
P_{\Lambda}(v,0) = 0\,,
\label{P0}
\end{equation}
in the range $0<v<1$.
Solutions of equation (\ref{P0}) for all Archimedean and Laves lattices have been studied in \cite{Codello2010}.
Equivalently, we can determine the critical $v$ as the inverse of the largest positive real eigenvalue of $\mathbb{W}_{\Lambda}(0)$.
This characterisation is very useful when the computation of the characteristic polynomial (\ref{P0}) becomes too demanding.
Near the critical point, and in terms of the reduced temperature $t=T/T_c-1$, the
critical exponent $\alpha$ is defined by the scaling $f_\Lambda(t) \sim
t^{2-\alpha}$. In particular, a logarithmic singularity of the free
energy, as the one present in (\ref{logdet}), is encoded in $\alpha=0$.

\paragraph*{Correlation length.}
The correlation length can be computed from the knowledge of the lattice mass since  $\xi_\Lambda = 1/m_\Lambda$.
This last is defined by the following small momenta expansion of the determinant,
\begin{equation}
P_\Lambda(v,\mathbf{k}) = Z_\Lambda(v)\, \Big[ m^2_\Lambda(v) + k^2 + O(k^4) \Big] \,,
\end{equation}
where $Z_\Lambda(v) \equiv \left. \frac{\partial }{\partial k^2} P_\Lambda(v,\mathbf{k}) \right|_{k=0}$ is the wave function renormalization.
The lattice mass can then be written as $m^2_\Lambda(v) = P_\Lambda(v,0)/Z_\Lambda(v)$ and the correlation length takes the form
\begin{equation}
\xi_\Lambda(v) = \sqrt{\frac{\left. \frac{\partial }{\partial k^2} P_\Lambda(v,\mathbf{k}) \right|_{k=0}}{P_\Lambda(v,0)}}\,.
%\sqrt{\frac{P_\Lambda\left(v,\frac{\pi }{2},\frac{\pi }{2}\right)-P_\Lambda(v,0,0)}{P_\Lambda(v,0,0)}}\,.
\label{cl}
\end{equation}
The correlation length critical exponent is defined by the relation $\xi_\Lambda \sim |t|^{-\nu}$ valid in the scaling region near the phase transition.

%
%The correlation length can also be computed by evaluating the
%determinant at a fixed value of the momenta, but to be able to do this one needs to know the momentum structure exactly, as in the case of the gaskets.

\section{Fractals}

\paragraph*{Approximate solutions on fractals.}
We now want to use our ability of solving the Ising model on an arbitrary periodic lattice to find approximations to the same problem but on fractal lattices of fractal dimension below two.
 
We will study Sierpinski carpets that are defined in an iterative
way.
%, where at each finite iteration the fractal is approximated by a periodic lattices where th
Let us consider a two dimensional $L\times L$ tile, where some
of the squares have been removed: we call this the generator of the
fractal. It is also the first iteration, $k=1$, of the sequence that
defines the fractal. Then, given the sequence at iteration
$k$, the next iteration, $k+1$, can be constructed by replacing every
existing square at the $k^{\rm{th}}$ iteration  with the
generator. In the limit $k\to\infty$ this defines a fractal. 

\begin{figure}%[ht!]
\begin{center}
\includegraphics[scale=0.6]{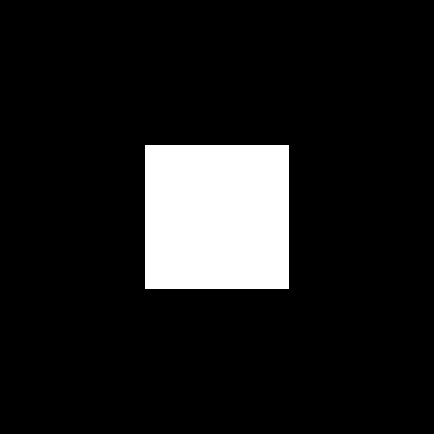}\quad\includegraphics[scale=0.6]{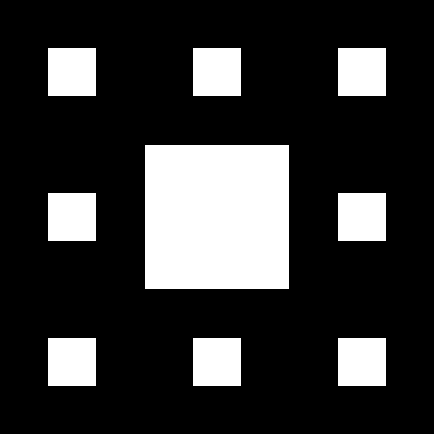}\quad\includegraphics[scale=0.6]{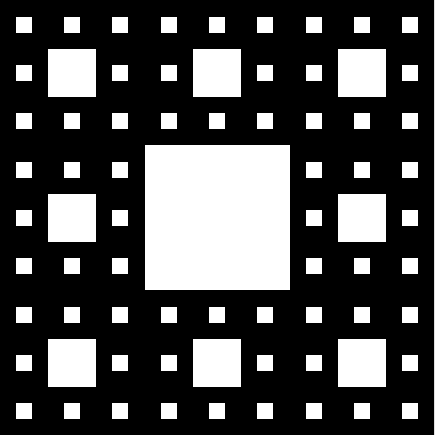}\quad\includegraphics[scale=0.6]{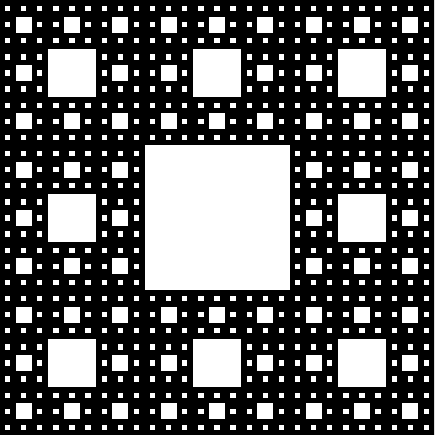}
\caption{ Illustration of ${\rm sc}(3,1)_k$ for $k=1,2,3,4$.  \label{fig:our_friend}}
\end{center}
\end{figure}

In our approach, the lattice $\Lambda_k$ at iteration $k$ is defined 
as the infinite repetition of the tiling obtained at level $k$. We assume the Ising model on the limiting 
fractal $\Lambda_\infty =  \lim_{k \to \infty} \Lambda_k$ exhibits the same behaviour
as  the Ising model defined on the fractal. In other words, we approximate the fractal's determinant $P_{\Lambda_\infty}(v,\mathbf{k})$  using periodic approximations
\begin{equation}
P_{\Lambda_\infty}(v,\mathbf{k}) = \lim_{k \to \infty} P_{\Lambda_k}(v,\mathbf{k}) \,.
\label{limit}
\end{equation}
%
%since we are in principle able to construct the polynomials
%$P_k(v,\mathbf{k})$ explicitly, at least for small $k$.
Furthermore we define  ${\rm sc}(L,b)_1$ to be the generator
where from $L\times L$ tile a $b\times b$ square is removed from
the center. Then ${\rm sc}(L,b)_k$ denotes the tiling at iteration
$k$. For illustration see Figure~\ref{fig:our_friend}. We will denote
with ${\rm SC}(L,b)_k$ the lattice obtained by tessellation of the plane with tile ${\rm sc}(L,b)_k$. 
We will also study Sierpinski gaskets. Theirs  generators are a $L\times L$
tile where a single $(L-1)\times(L-1)$ block has been
removed. Sierpinski gaskets have a finite ramification number, i.e. one
can remove arbitrarily large pieces by cutting a finite number of links.

We denote with $T_{k}$ the critical temperature of the lattice $\Lambda_k$ where
$\Lambda$ is understood from the context and  similarly  for the
correlation length $\xi_k$.

%Sierpinski carpet(s), i.e. all fractal constructed by eliminating
%sites in a $L \times L$ square lattice,
%, which we solve exactly at any finite iteration $k$

% \paragraph*{Fourier expansion.}
% %
% On a square lattice we can expand (\ref{P}) in a Fourier series:
% %
% \begin{equation}
% P_n(v,\mathbf{k}) = \sum_{\mathbf{m}} P_n(v,\mathbf{m}) \cos m_x k_x \cos m_y k_y \,,
% \label{PFourier}
% \end{equation}
% %
% where all the $P_n(v,\mathbf{m})$ are in principle known exactly. Realistically their analytical form becomes an intractable function of $v$ very soon, but for fixed $v$ they are just numbers and as such can be determined precisely.
% \\
% \\
% Explain notation and make a figure!

%%%%%%%%%%%%%%%%%%%%%%%
\subsection{Sierpinski carpets} \label{sec:sierpinski_carpet}
%%%%%%%%%%%%%%%%%%%%%%%

\paragraph*{Explicit form of $\mathbb{W}$.}

\begin{figure}%[ht!]
\begin{center}
\includegraphics[scale=0.35]{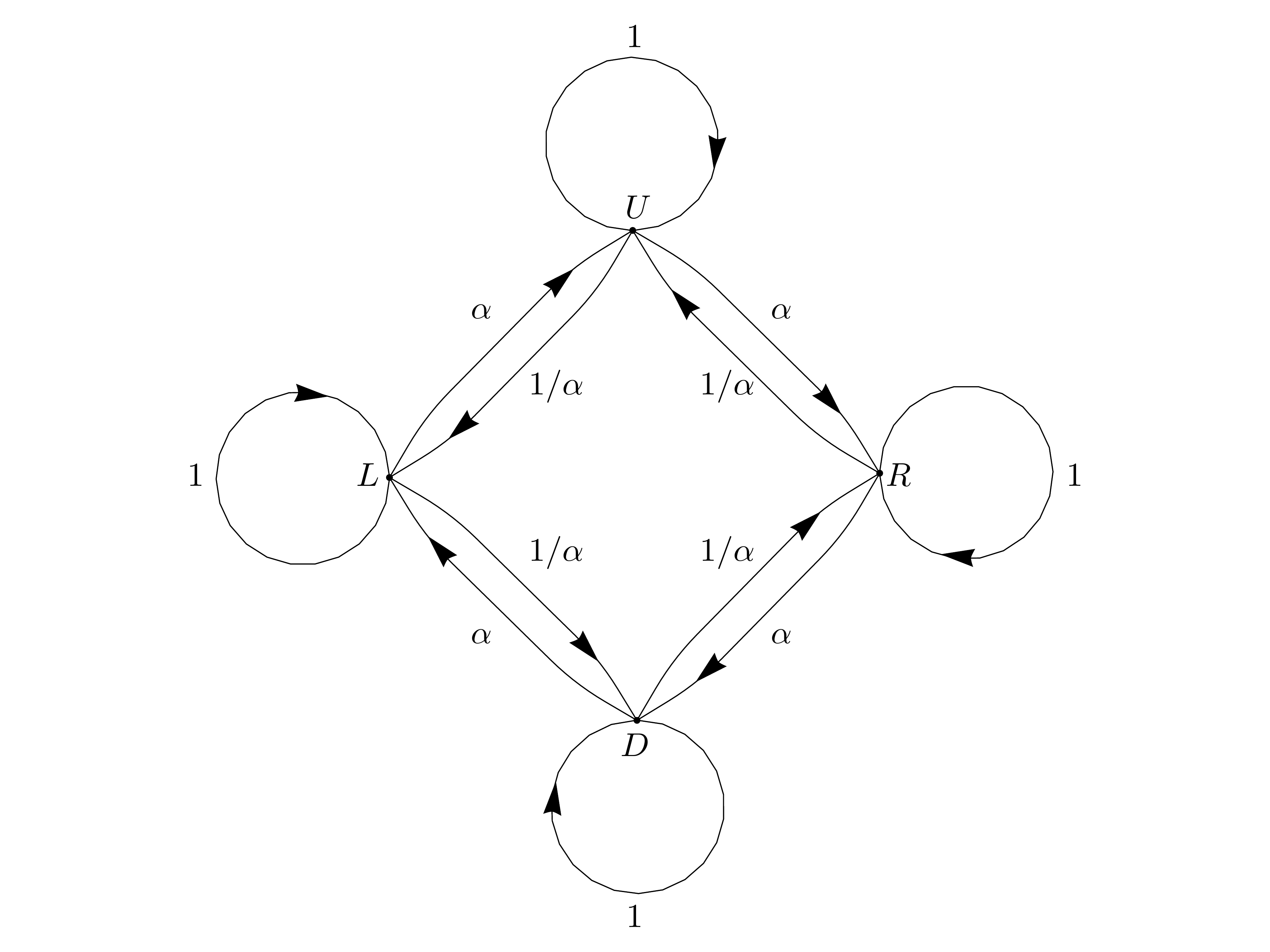}
\caption{Graph representing the transition matrix $\mathbb{W}_{Ising}(0,0)$ for the standard Ising model on a square lattice.}
\label{graph}
\end{center}
\end{figure}

We can construct the transition matrix $\mathbb{W}_{\Lambda_k}(k_x,k_y)$ for
any finite iteration Sierpinski carpet $\Lambda_k$ exactly. Explicitly, this
can be constructed as the adjacency matrix of a weighted directed graph, in
which each node of the graph represents one of the four directions
$U,L,D,R$ associated to each site of the generator of $\Lambda_k$.
In particular, when $k_x=k_y=0$ we can represent $\mathbb{W}_{\Lambda_k}(0,0)$ as the graph

\begin{align}
U_{i,j} &\overset{ w_{i,j}}{\longrightarrow} U_{i-1,j} & U_{i,j}
&\overset{w_{i,j}/\alpha}{\longrightarrow} L_{i-1,j} & U_{i,j}
&\overset{\alpha w_{i,j}}{\longrightarrow} R_{i-1,j}     \nonumber\\
L_{i,j} &\overset{\alpha w_{i,j}}{\longrightarrow} U_{i,j-1} & L_{i,j} &\overset{w_{i,j}}{\longrightarrow} L_{i,j-1} & L_{i,j} &\overset{w_{i,j}/\alpha}{\longrightarrow} D_{i,j-1}      \nonumber\\
D_{i,j} &\overset{\alpha w_{i,j}}{\longrightarrow} L_{i+1,j} & D_{i,j}
&\overset{w_{i,j}}{\longrightarrow} D_{i+1,j} & D_{i,j}
&\overset{w_{i,j}/\alpha}{\longrightarrow} R_{i+1,j}    \nonumber\\
R_{i,j} &\overset{w_{i,j}/\alpha}{\longrightarrow} U_{i,j+1} & R_{i,j}&\overset{\alpha w_{i,j}}{\longrightarrow} D_{i,j+1} & R_{i,j} &\overset{w_{i,j}}{\longrightarrow} R_{i,j+1} \nonumber
\label{matrix}
\end{align}
%
%\end{widetext}
%
where the indices are modulo $L$ and $\alpha = e^{i\frac{\pi}{4}}$ is
the complex amplitude required by the Feynman-Vodvicenko method. Basically, in terms of directions,  clockwise arrows have  amplitude
$\alpha$,  counter clockwise arrows have amplitude $\alpha^{-1}$, while
self connections have amplitude one. The
weights $w_{i,j}$ are chosen equal to the matrix representation of the
generator of the lattice $\Lambda_k$, setting $w_{i,j}=1$ if the
site $(i,j)$ of the generator exists and  $w_{i,j}=0$ if it is depleted.
 The case of the standard Ising
model on square lattice can be represented as in Figure~\ref{graph}. The momentum dependence of
$\mathbb{W}_{\Lambda_k}(k_x,k_y)$ is
obtained by multiplying the links outgoing from $U$ with $e^{ik_y}$,
from $L$ with  $e^{ik_x}$, from $D$ with  $e^{-ik_y}$, and from $R$ with  $e^{-ik_x}$.

Although we are here interested in approximating fractals, by properly
choosing $w_{i,j}$ this construction gives the transition matrix for
any lattice with rectangular tile. For example, our construction
encompasses the exact solution of the Ising model on all possible two
dimensional depleted lattices, including the case of a random basic tile.

\paragraph*{Standard Ising model.}

%%%%%%%%%%%%%%%%%%%
%%%%%%%%%%%%%%%%%%%%%%

We shortly review the solution of the standard two dimensional Ising model to exemplify the method.
From Figure~\ref{graph} we immediately reconstruct the transition matrix
\begin{eqnarray}
\mathbb{W}_{Ising} (k_x,k_y)=
\left(
\begin{array}{cccc}
 e^{i k_y} & \frac{1}{\alpha }e^{i k_y} &
   0 & \alpha e^{i k_y}    \\
 \alpha e^{i k_x}   & e^{i k_x} & \frac{1}{\alpha
   }e^{i k_x}& 0  \\
0 &\alpha e^{-i k_y}   & e^{-i k_y} & \frac{1}{\alpha } e^{-i k_y}\\
 \frac{1}{\alpha } e^{-i k_x}& 0&\alpha e^{-i k_x} & e^{-i k_x} 
     \\
\end{array}
\right)\,,
\end{eqnarray}
with $\alpha = e^{i \frac{\pi}{4}}$. The determinant is readily computed and gives the well known Onsager's solution \cite{Onsager_1944}:
\begin{eqnarray}
P_{Ising} (v,\mathbf{k}) = \left(1+v^{2}\right)^2-2 \,v  \left(1-v^{2}\right)\Big(\!\cos k_x+ \cos k_y \Big)\,.
\end{eqnarray}
Setting $P_{Ising} (v,0) =(1-2v - v^2)^2 = 0$ gives $v_c=0.414214...$ as the only solution in the range $0<v<1$; this correspond to the critical temperature $T_c = 2.26919...$ first computed by Kramer and Wannier using duality arguments \cite{Kramers_Wannier_1941}. Finally, using (\ref{cl}) we find the exact form for the correlation length
\begin{eqnarray}
\xi_{Ising}(v) = 2 \sqrt{\frac{v-v^3}{\Big(1-2v - v^2\Big)^2}}\,,
\end{eqnarray}
which clearly diverges for $T\sim T_c$ since the denominator vanishes.

To check our formalism we can try to solve a redundant version of the
two dimensional Ising model, defined on a tile of size $L>1$. With our
previously defined notations, they are equivalent to ${\rm SC}(L,0)_1$.
For example in the cases $L=2,3$ we find to following characteristic polynomials,
\begin{eqnarray}
P_{\rm{SC}(2,0)_1} (v,0) &=& (1+v^2 )^4 (1+2v-v^2 )^2 (1-2v - v^2 )^2 \nonumber \\
P_{\rm{SC}(3,0)_1} (v,0) &=& (1 - 2 v - v^2)^2 (1 + 2 v + 2 v^2 - 2 v^3 + v^4)^4 (1 - v + 2 v^2 + v^3 + v^4)^4 \,,
\end{eqnarray}
which indeed have only the $v_c=0.414214...$ solution in the range $0<v<1$. This represents a non-trivial check of our ability to construct the transition matrices for an arbitrary lattice. The momentum dependence becomes rapidly very complicated but a similar analysis can be made for the correlation length.

%%%%%%%%%%%%%%%%%
\subsection{Analytical results} \label{sec:analytical_res}
%%%%%%%%%%%%%%%%%

\begin{figure}%[ht!]
\begin{center}
\includegraphics[scale=0.7]{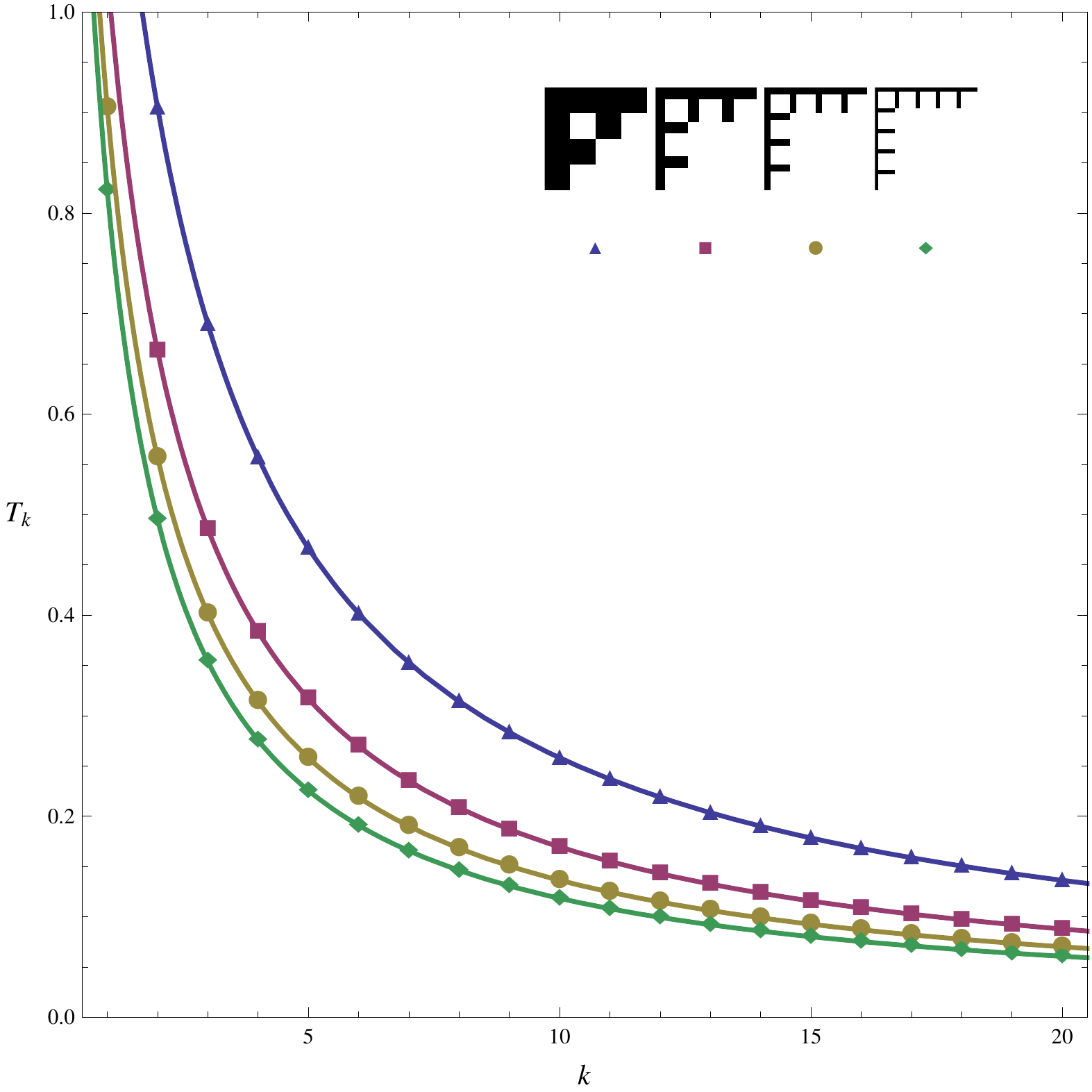}
\caption{The critical temperature for the Sierpinski gaskets as a function of $k$. Convergence towards $T_\infty=0$ is very slow (logarithmic).
The $k=3$ generators for (from left) the $L=2,3,4,5$ Sierpinski gaskets considered in the study are shown in the legend.}
\label{fig_Tc_sg}
\end{center}
\end{figure}

\paragraph*{Analytical solutions for the Sierpinski gaskets.}
Analytical relations can be found for all the Sierpinski gaskets defined on a $L\times L$ grid.
The cases $L=2,3,4,5,...$ are shown in the legend of Figure~\ref{fig_Tc_sg}.
We are able to give the analytical form for the determinant:
\begin{eqnarray}
%P_k (v,\mathbf{k}) = \left(1+v^{2^{k+1}}\right)^2-2 \,v^{2^k}  \left(1-v^{2^{k+1}}\right)\Big(\!\cos 2^k k_x+ \cos 2^k k_y \Big)\,.
P_{\Lambda_k} (v,\mathbf{k}) = \left(1+v^{2L^{k}}\right)^2-2 \,v^{L^k}  \left(1-v^{2L^{k}}\right)\Big(\!\cos L^k\, k_x+ \cos L^k\, k_y \Big)\,.
\label{sgf}
\end{eqnarray}
Since $v<1$ the infinite iteration limit leads just to one $\lim_{k \to \infty} P_{\Lambda_k} (v,\mathbf{k}) = 1$.
Thus $\Phi_{\Lambda_\infty}(v) = 1$ as in the one dimensional case and the singular non-trivial part of the free energy per spin is zero.
We recover in this way the result that Ising models on Sierpinski gaskets do not magnetise~\cite{gefen_1984,Burioni}.
%
%%%%%%%%%%%%%%%%%%%
%%%%%%%%%%%%%%%%%%%%%%

\begin{figure}%[ht!]
\begin{center}
\includegraphics[scale=1.2]{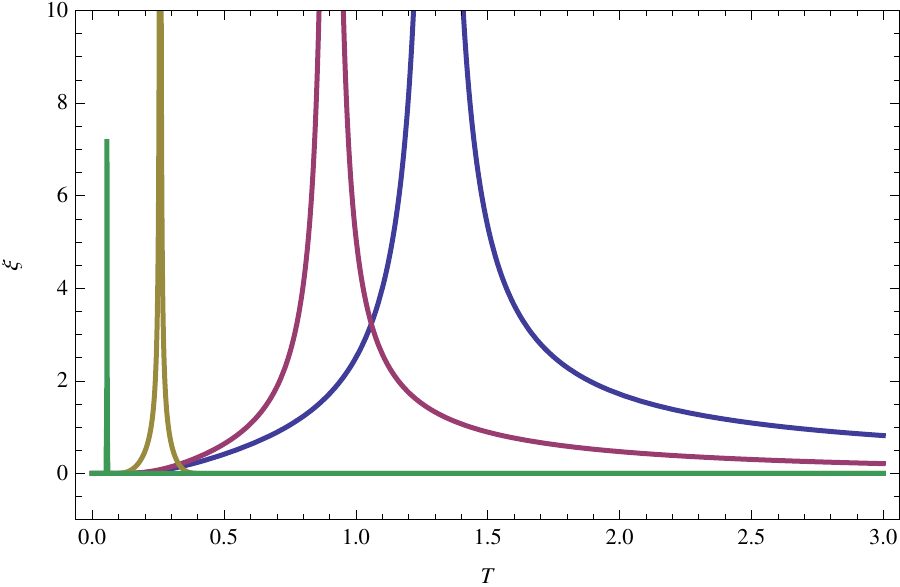}
\caption{The correlation length $\xi_k$ for the $L=3$ Sierpinski gasket as a function of $T$ for the values $k = 1,2,10,50$.}
\label{fig_xi_sg}
\end{center}
\end{figure}

Even if $T_\infty=0$, it is interesting to infer from (\ref{sgf}) the exact critical temperature for any finite $k$:
\begin{eqnarray}
%T_k = \frac{1}{\textrm{Arctanh}\left(\!\sqrt{2}-1\right)^{1/2^{k}}}\\
T_{k} = \frac{2}{\log \frac{1+\left(\!\sqrt{2}-1\right)^{1/L^k}}{
   1-\left(\!\sqrt{2}-1\right)^{1/L^k}}}\,.
\label{sgT}
\end{eqnarray}
This relation is instructive since it shows that convergence to the limiting value is very slow, more precisely logarithmic, as can be seen in Figure~\ref{fig_Tc_sg}.
It is interesting to look also at the correlation length, which from Eq.~(\ref{cl}) turns out to be
\begin{eqnarray}
%\xi_k (v) = 2 \left[ \frac{v^{2^k}-v^{3\times 2^k}}{\left(v^{2^k} \left(v^{2^k}+2\right)-1\right)^2} \right]^\frac{1}{2} \,.
\xi_k (v) = 2 \sqrt{ \frac{v^{L^k}-v^{3L^k}}{\left(v^{L^k} \left(v^{L^k}+2\right)-1\right)^2} }\,.
\label{sgxi}
\end{eqnarray}
This relation is visualised in Figure~\ref{fig_xi_sg}. The correlation length critical exponent is one, as in the one dimensional case, for all Sierpinski gaskets.

%%%%%%%%%%%%%%%%%%%%%%%%%%%%
%\begin{figure}
%\label{fig_sgaskets}
%\begin{centering}
%\includegraphics{sgaskets.pdf}
%%\par
%\end{centering}
%\caption{The Sierpinski gaskets considered in the study at their $k=3$ approximation for (from left) $L=2,3,4,5$.}
%\end{figure}
%% table / numerical requirements.
%%%%%%%%%%%%%%%%%%%%%%%%%%%

%Similar analytical relations can be found for all the other Sierpinski gaskets with $L=3,4,...$ and the generalisation of (\ref{sgf}) is:
%%
%\begin{eqnarray}
%P_k (v,\mathbf{k}) = \left(1+v^{2L^{k}}\right)^2-2 \,v^{L^k}  \left(1-v^{2L^{k}}\right)\Big(\!\cos L^k\, k_x+ \cos L^k\, k_y \Big)\,.
%\label{sgf}
%\end{eqnarray}
%
%The exact critical temperature for any $k$ to be:
%%
%\begin{eqnarray}
%%T_k = \frac{1}{\textrm{Arctanh}\left(\!\sqrt{2}-1\right)^{1/2^{k}}}\\
%T_k = \frac{2}{\log \frac{1+\left(\!\sqrt{2}-1\right)^{1/L^k}}{
%   1-\left(\!\sqrt{2}-1\right)^{1/L^k}}}\,.
%\label{sgTL}
%\end{eqnarray}
%
%The correlation length becomes:
%%
%\begin{eqnarray}
%\xi_k (v) = 2 \left[  \frac{v^{L^k}-v^{3L^k}}{\left(v^{L^k} \left(v^{L^k}+2\right)-1\right)^2} \right]^\frac{1}{2}\,.
%\label{sgxiL}
%\end{eqnarray}

It is clear that a similar analysis, with similar conclusions, can be made for other families of fractals with finite ramification number. It is probably possible to obtain a closed formula for $P_k(v,\mathbf{k})$ for any fractal with this property. This is a clear indication of their triviality and effective one dimensional behaviour.

%%%%%%%%%%%%%%%%%%%
%%%%%%%%%%%%%%%%%%%%%%

\paragraph*{Analytical solution of the $k=1$ Sierpinski carpet.}
We can give the analytical solution of the Sierpinski carpet $\rm{SC}(3,1)_1$.
The critical temperature is the solution of 
\begin{eqnarray}
P_{\rm{SC}(3,1)_1}(v,0)  = \Big(1 - 4 v^3 + 5 v^4 - 16 v^5 - 10 v^6 - 20 v^7 + v^8 - 24 v^9 + 2 v^{10} + v^{12}\Big)^2\,.
\label{pol}
\end{eqnarray}
The only root in the range $0<v<1$ is $v_c = 0.495968...$ which gives $T_c = 1.83842...$ as reported in the $k=1$ entry of Table I.
Note that it is a non-trivial fact and a consistency check that the 12$^{th}$ degree polynomial in Eq.~(\ref{pol}) has only one real solution in the physical range.

In this case we are also able to determine the full momentum dependence of the determinant
\begin{eqnarray}
&& P_{\rm{SC}(3,1)_1}(v,\mathbf{k}) = \qquad \nonumber\\
&& \quad 1+10 v^4 + 20 v^6 + 119 v^8 + 324 v^{10} + 876 v^{12} + 1284 v^{14} +983 v^{16} + 412 v^{18} + 58 v^{20} + 8 v^{22} + v^{24} \nonumber\\
&& \quad -4 v^3 (1-v^2)^2 \Big(1 + 6 v^2 + 21 v^4 + 52 v^6 + 69 v^8 + 72 v^{10} + 29 v^{12} + 6 v^{14} \Big)  \Big(\! \cos 3 k_x + \cos 3 k_y  \Big) \nonumber\\
&& \quad -2 v^6 (1-v^2)^4 \left(7 + 18 v^2 + 24 v^4 + 14 v^6 + v^8 \right) \Big(\!\cos 3(k_x+k_y) + \cos 3(k_x-k_y) \Big) \nonumber\\
&& \quad +2 v^6 (1-v^2)^5 \left(1 + 4 v^2 + 3 v^4\right)  \Big(\! \cos 6 k_x + \cos 6 k_y \Big) \,.
\label{detk1}
\end{eqnarray}
This relation clearly illustrates how non-trivial are the explicit
solutions already at the level of the first iteration. It also shows
how higher harmonics are excited, and that in
$P_{\Lambda_k}(v,\mathbf{k})$ the coefficients of the trigonometric
functions are polynomials in $v$.
We have obtained similar relations for many  $k=1$
non-trivial fractals, i.e. with infinite ramification number, while we have not been able to obtain closed analytical forms for the determinants as a function of $k$, and we suspect this to be a formidable task, even if not hopeless. Such a closed formula will constitute an explicit exact solution of the model.
\newpage
Finally, we also report the correlation length in the $k=1$ case. Inserting Eq.~(\ref{detk1}) into Eq.~(\ref{cl}) gives
{\tiny
\begin{equation}
\xi_1(v) = 2 \left( \frac{v^3 \left(v^2-1\right)^2 \left(v^{15}+12 v^{14}+18 v^{13}+58
   	v^{12}-13 v^{11}+144 v^{10}-20 v^9+138 v^8+7 v^7+104 v^6+2 v^5+42 v^4+5
   	v^3+12 v^2+2\right)}{\left(v^{12}+2 v^{10}-24 v^9+v^8-20 v^7-10 v^6-16 v^5+5
   	v^4-4 v^3+1\right)^2} \right)^\frac{1}{2} \,. \nonumber
\end{equation}
}
This correlation length diverges consistently at $v_c = 0.495968...$ and when expressed in terms of the reduced temperature is plotted as the upper curve in Figure~\ref{fig:xi_vs_t}. Clearly $\nu=1$ as expected from universality.

%%%%%%%%%%%%%%%%%%%%%%%%%%%
\begin{figure}
\begin{center}
\includegraphics[scale=1]{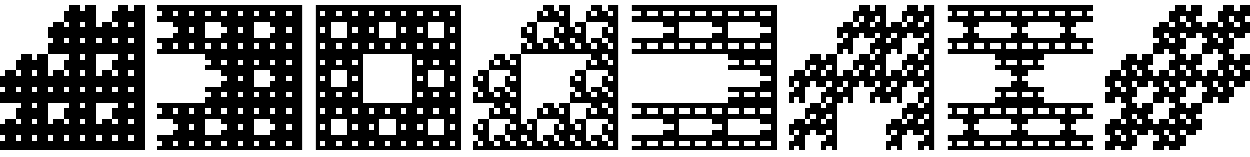}
\label{fig_sc}
\caption{The $L=3$ Sierpinski carpet(s) considered in the study at their $k=4$ approximation.}
\end{center}
\end{figure}
% table / numerical requirements.
%%%%%%%%%%%%%%%%%%%%%%%%%%

%%%%%%%%%%%%%%%%%
\subsection{Critical temperatures}
%%%%%%%%%%%%%%%%%

\paragraph*{Numerical analysis of critical temperatures.}
% #summary of the numerical problem,  size of matrix, references to eq 
% #k= 1 case : analyical expression of the characteristic poly =>
% numerical of estimation of all the roots
% #k =2  case : -Algorithm to go beyond : Shifted Arnoldi.
%              -knowing k=1 -> to get k > 1
% #stability : why is it numerically stable, and why are we convinced
% that we get the correct eigenvalue.
%
% #table results & numerical requirements/limitation.
%
We have reduced the calculation of the critical temperature on a
lattice $\Lambda_k$ to finding the largest positive real eigenvalue
$\lambda_{\Lambda_k}$ of
the matrix $\mathbb{W}_{\Lambda_k}(0,0)$ corresponding to the 
weighted adjacency graph defined in the
previous section. The
$4L^{2k}\times4L^{2k}$ matrix  $\mathbb{W}_{\Lambda_k}$ is sparse, but its size
grows rapidly as a function of the $k$. For $k>1$ we are not able to
calculate analytically the eigenvalues but we instead resort to
numerical calculations.  

% k >  2 case
We use the shifted Arnoldi solver of Mathematica, which uses the ARPACK
library and is sufficient for an initial proof of concept. The
algorithm can be used to compute an arbitrary number of eigenvalues
in the neighbourhood of a complex number, usually referred as the
shift parameter. In our study, we compute the eigenvalue
$\lambda_{\Lambda_k}$ using as a shift the eigenvalue
$\lambda_{\Lambda_{k-1}}$.

% stability
% - no guarantee that there is no complex eigenvalue close to the real
% axis -> are there other eigenvalue around ? 
% two practical solution : exact spectrum for k=3  and investigation
% of the stability of the results vs the value of the shift. 

\begin{figure}
\begin{centering}
\includegraphics[width=0.8\textwidth]{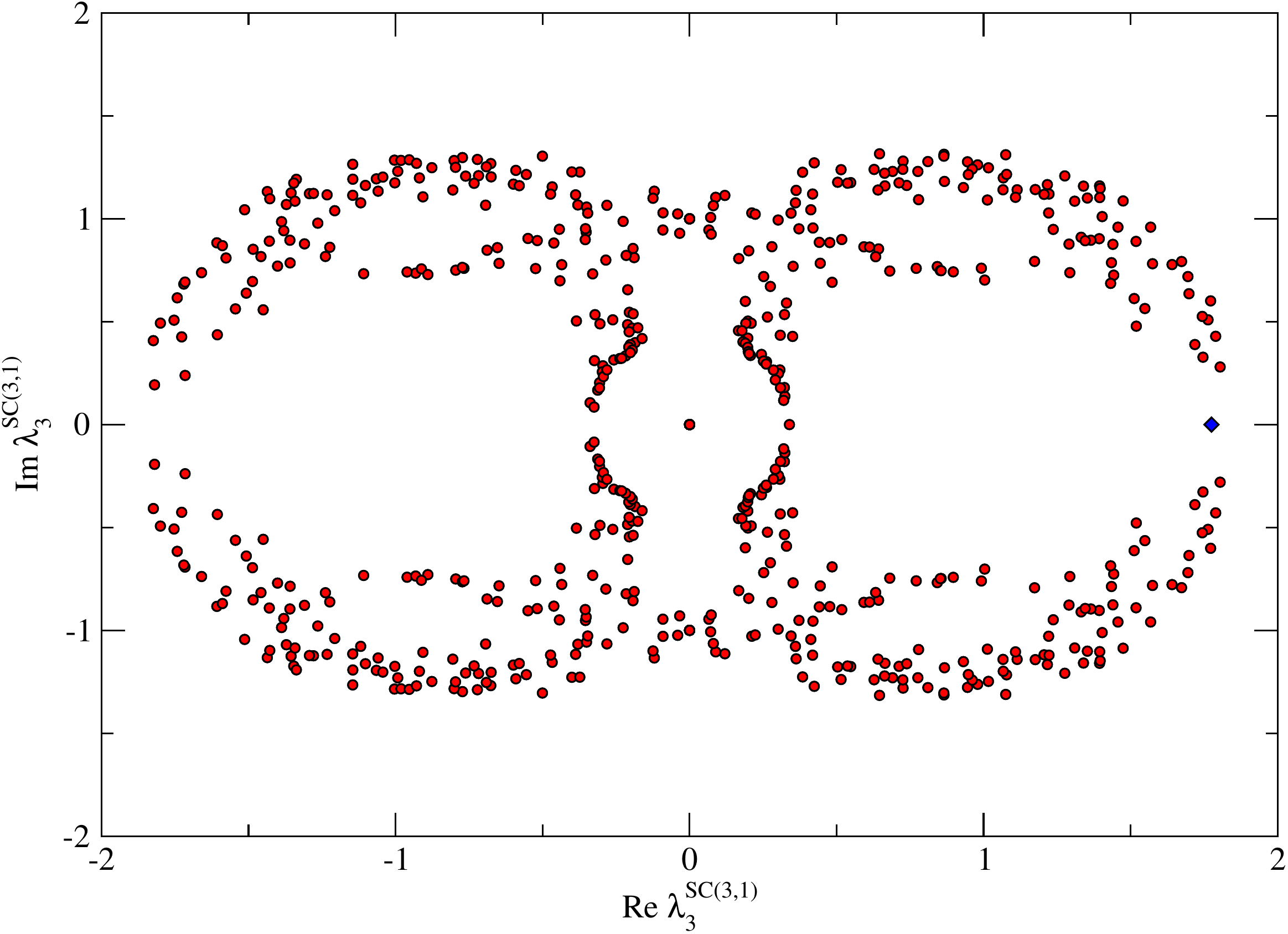}
\par
\end{centering}
\caption{Eigenvalues of $\mathbb{W}_{\rm{SC}(3,1)_3}$. The only real eigenvalue 
larger than one is shown using a blue diamond.\label{fig:spec_k3}}
\end{figure}

A caveat of  our approach is that it would give a wrong
estimate of $\lambda_{\Lambda_k}$ if there was a complex eigenvalue
with a small imaginary part in the neighborhood of
$\lambda_{\Lambda_{k-1}}$. However, it turns out that the eigenvalue is isolated, as for instance
illustrated in Figure~\ref{fig:spec_k3} in the case of $\rm{SC}(3,1)_3$, for which we can
compute the entire spectrum of the matrix $\mathbb{W}_{\rm{SC}(3,1)_3}(0,0)$. For larger $k$ we investigate the stability
of our prediction depending on the value of the chosen shift parameter. For all the
cases considered we observed that changing the shift leads to the same estimate of the critical temperature when we are able
to compute enough eigenvalues. If we could not compute enough
eigenvalues, then we find only non real eigenvalues. Therefore, in practice this numerical
limitation does not arise. 

%%%%%%%%%%%%%%%%%%%%%%%%%%%

%%%%%%%%%%%%%%%%%%%%%
\begin{figure}[h!]
\begin{centering}
\includegraphics[scale=0.7]{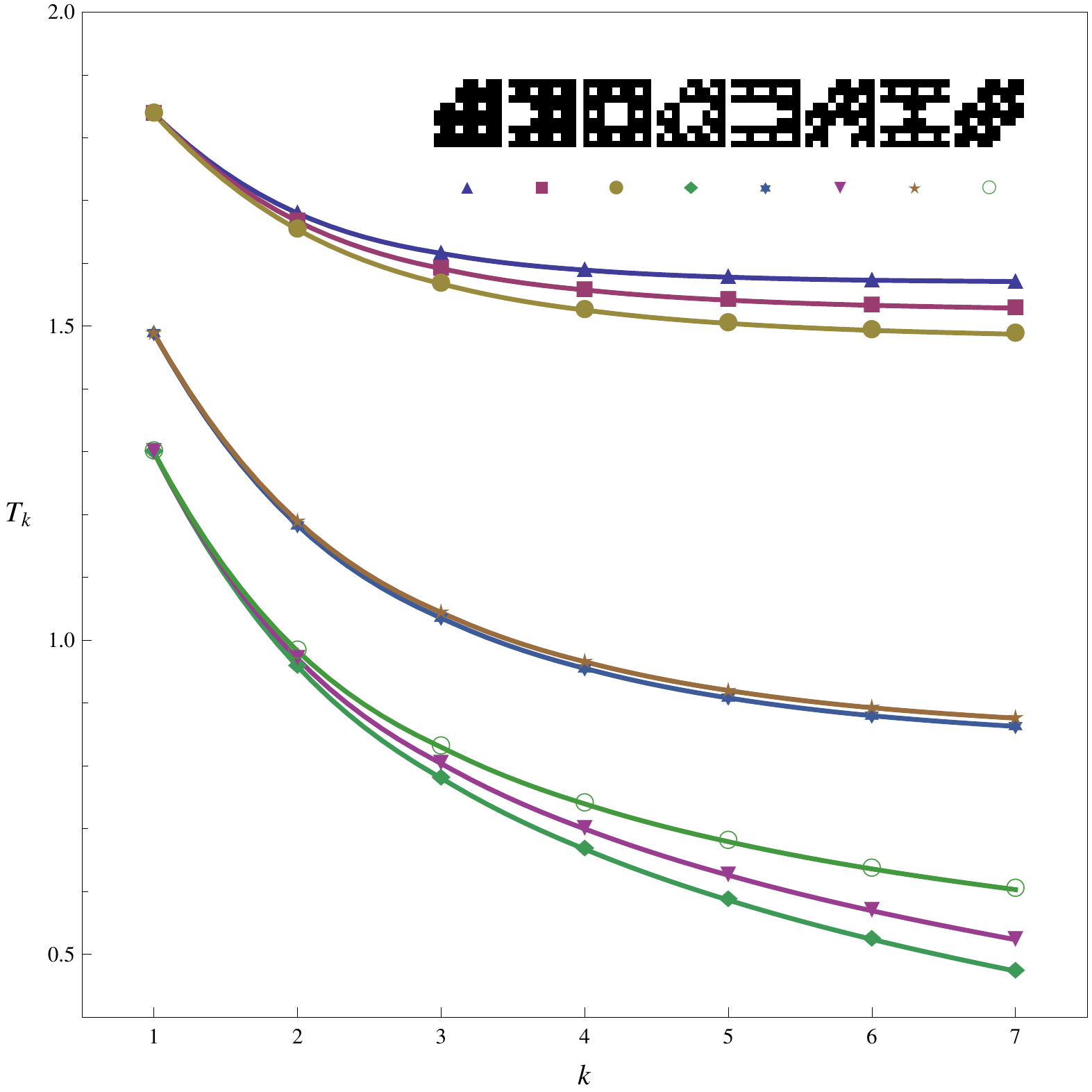}
\par
\end{centering}
\caption{Exact critical temperatures for the first seven approximands to the non-trivial $L=3$ Sierpinski carpets considered, together with a fit.\label{fig:L3_TC}}
\end{figure}
%%%%%%%%%%%%%%%%%%%%%

\begin{center}
\begin{table}[h!]
\begin{tabular}{ccccccccc}
\hline\hline
%&&\multicolumn{7}{c}{ $k$ } \\
Generator & $d_f$ & $k=1$  & $k=2$  & $k=3$  & $k=4$  & $k=5$  & $k=6$ & $k=7$     \tabularnewline
\hline 
$ \includegraphics[scale=0.15]{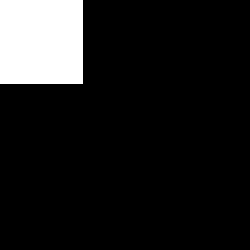} $ & $1.893$ &$1.83842$ &$1.67971$& $1.61601$& $1.58935$ &$1.57798$& $1.57310$& $1.57099$\tabularnewline
$ \includegraphics[scale=0.15]{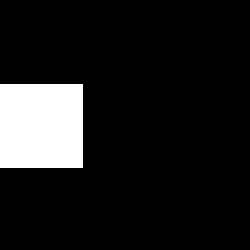}$  &  $1.893$  &$1.83842$ &$1.66680$& $1.59188$& $1.55769$ &$1.54140$& $1.53319$& $1.52872$\tabularnewline
$ \includegraphics[scale=0.15]{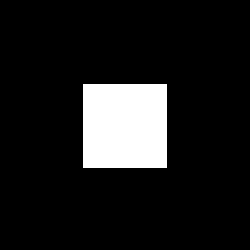}$ &  $1.893$ &$1.83842$ &$1.65386$& $1.56759$& $1.52566$ &$1.50446$& $1.49331$& $1.48719$\tabularnewline
$ \includegraphics[scale=0.15]{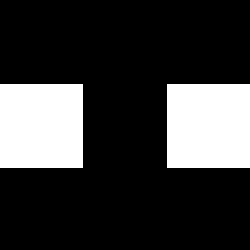}$ & $1.771$ &$1.48866$ &$1.18962$& $1.04440$& $0.965875$ &$0.920115$& $0.892608$& $0.875999$\tabularnewline
$ \includegraphics[scale=0.15]{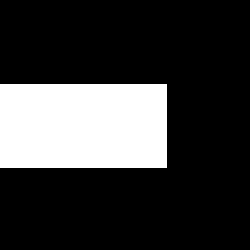}$ & $1.771$ &$1.48866$ &$1.18310$& $1.03567$& $0.955384$ &$0.908286$& $0.879960$& $0.862996$\tabularnewline
$ \includegraphics[scale=0.15]{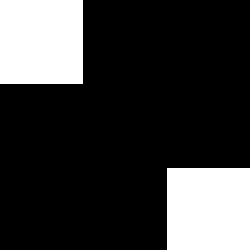}$ & $1.771$ &$1.29944$ &$0.983021$& $0.830078$& $0.739657$ &$0.679436$& $0.636087$&$0.603146$\tabularnewline
$ \includegraphics[scale=0.15]{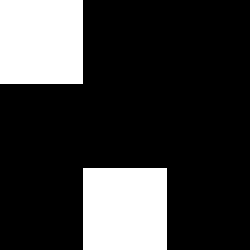}$  & $1.771$ &$1.29944$ &$0.97052$& $0.80394$& $0.699862$ &$0.626109$& $0.569437$& $0.523428$\tabularnewline
$ \includegraphics[scale=0.15]{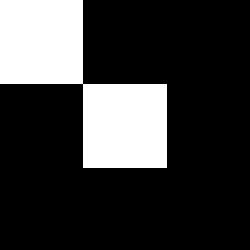}$  & $1.771$ &$1.29944$ &$0.958433$& $0.780739$& $0.667582$ &$0.586519$& $0.524012$& $0.473526$\tabularnewline
\hline 
\hline
\end{tabular}
\caption{Exact critical temperatures for the various iterations of the
  $L=3$ Sierpinski carpets. The fractals are ordered from the
  largest to the lowest critical temperature. Note that for $k=1$
  different lattices have the same critical temperature, since at
  this level, there are exactly the same.\label{tab:SierpinskiTc}}
\end{table}
\end{center}
%%%%%%%%%%%%%%%%%%%%%%%%%%%%%%%%%

Using this procedure, we can
calculate the critical temperature up to $k=7$, for the $L=3$
fractals. To achieve this, we need to find a specific eigenvalue of a
$19~131~876\times19~131~876$ matrix. The calculations are limited by
the available memory. All the computations performed in this section
have been achieved running Mathematica on a single node with 20 cores
and 128GB of memory. Without further method improvements a machine
with more memory would be needed to compute $T_k$ for $k>7$.

The results for the $L=3$ Sierpinski carpets are given in
Table~\ref{tab:SierpinskiTc}.  They are illustrated for the 8 fractals
considered in Figure~\ref{fig:L3_TC}.
%
% comments on the results L=3
% the closest form d=2 tbe closest from the ising Tc
% extraploation  : why we don't it : scaling relation ??
The fractals considered here have two different fractal
dimensions. However they differ by there number of active bonds. The
three fractals that have a dimension close to two show a much faster
convergence than the others. As seen in the previous section, in the
case of the Sierpinski gaskets, fractals
can show a very slow convergence in the $k\to\infty$ limit. 
We thus see that, for the fractals considered, the smaller the fractal dimension, the slower is the pace of convergence.
We do not attempt to perform any extrapolation to
$k\to\infty$ since, in our set-up, we lack a scaling theory that governs this limit as the volume is already
infinite for every $k$.
% Therefore we do not attempt to perform any extrapolation to
%$k=\infty$. 

\begin{figure}
\begin{centering}
\includegraphics[width=0.25\textwidth]{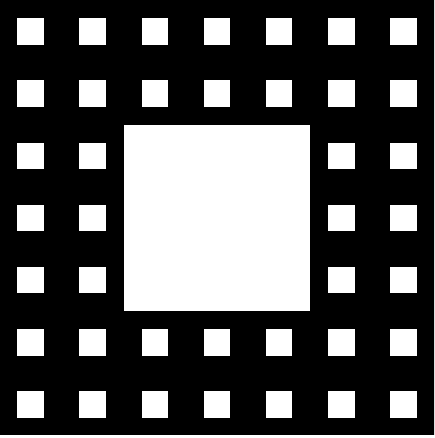}\qquad\includegraphics[width=0.25\textwidth]{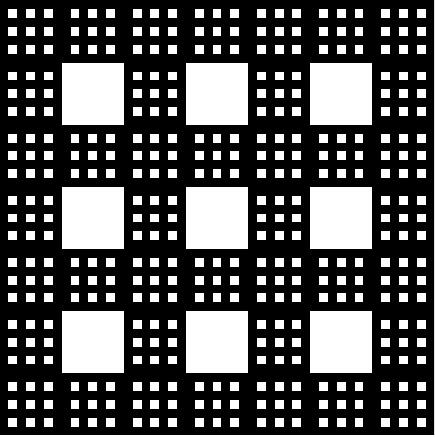}
\par
\end{centering}
\caption{Generators $\rm{sc}(7,3)_{2}$ (left) and
  $\widetilde{\rm{sc}}(7,3)_2$   (right).\label{fig:SC73}}
\end{figure}

As mentioned earlier, the method can be easily applied to other fractals,
and we investigate some of them with  $L=4,5$ and $7$ as summarized in
Table~\ref{tab:OtherSierpinskiTc}.  The fractal
$\widetilde{\rm{SC}}(7,3)$ is defined by removing nine distinct uniformly
distributed cells as illustrated in Figure~\ref{fig:SC73}. The two
fractals with $L=7$ generators considered here have the same fractal
dimension but different lacunarity. They have been discuss in \cite{gefen:1980aa,gefen:1984aa}.

% comments on the results for other L
% As pointed out by \cite{}, the two
% $L=7$ fractals have different lacunarity. Our result indicates that
% they seem to convergent to very different critical temperature.
% notes 
% 980218 : "one can decide whether a lattice is able to support an
% order-disorder transition of the Ising type by looking at its order
% of ramification R, which is finite if after eliminating a finite
% number of bonds one can isolate an arbitrarily large
% sublattice. Only if R = Ã¢ÂÂ a phase transition occurs."

%%%%%%%%%%%%%%%%%%%%%%
\begin{center}
\begin{table}[t]
\begin{centering}
 \begin{tabular}{ccccccc}
\hline\hline
Fractal & $d_f$  & $k=1$  & $k=2$  & $k=3$  & $k=4$  & $k=5$       \tabularnewline
\hline 
$\rm{SC}(4,2)$ & $1.792$ & $1.62129$ & $1.36891$ & $1.25015$ &
$1.18451$  & $1.14280$ \tabularnewline %  (ix=211)
$\rm{SC}(5,1)$    & $1.975$ & $2.11926$ & $2.07899$& $2.06904$
&$2.06672$  & - \tabularnewline % (ix=1)
$\rm{SC}(5,3)$   & $1.723$ & $1.48748$ & $1.19857$ & $1.05787$  &
$0.97483$  & - \tabularnewline % (ix=2)
$\rm{SC}(7,3)$    & $1.896$ & $1.92863$ & $1.85117$& $1.8334$ & $1.82927$  & - \tabularnewline % (ix=1)
$\widetilde{\rm{SC}}(7,3)$&  $1.896$  & $1.57100$& $1.39728$ & $1.34601$  &
$1.32719$ & - \tabularnewline %  (ix=2)
\hline\hline
\end{tabular}
\par\end{centering}
\caption{Exact critical temperatures for the various Sierpinski carpets
  with $L>3$.\label{tab:OtherSierpinskiTc}}
\end{table}
\end{center}
%\end{widetext}
%%%%%%%%%%%%%%%%%%%%%%%%

\paragraph*{Comparison with Monte Carlo approach.}

Defining $\Lambda_{k,l}\equiv {\rm sc}(L,b)_{k,l}$ as a finite lattice defined by an
$l\times l$  array of ${\rm sc}(L,b)_k$ building blocks.  Defining
$T_{k,l}$ as the critical temperature of the lattice $\Lambda_{k,l}$, 
where obviously for a finite $l$ the critical temperature is defined  for
instance as the maximum of the specific heat. Usually Monte Carlo
studies reports $T_{k,1}$ whereas we compute $T_{k,\infty} =
\lim_{l\to \infty} T_{k,l}$. While $T_{k,l}$ depends on the lattice
definition  of the critical temperature,  the value of
$T_{k,\infty}$ is unique.  The critical temperature of the fractal is approached
in the limit of $k\to\infty $ in both cases. Since it is proven that
for fractals with infinite ramification number $T_{\infty} > 0$, the
two approaches must yield to the same limiting value. The rate of
convergence is a priori unknown in both cases.

In table~\ref{tab:comparison} we compare  our results for
$T_{k,\infty}$ with the results for $T_{k,1}$
obtained using Monte Carlo
simulations for various fractals. The lattice estimates of the critical temperature are in a
good agreement with our results and almost always within the estimated
errors. As expected from the previous considerations, the fractals with the highest fractal dimensions
exhibit also a better agreement.

%%%%%%%%%%%%%%%%%%%%%%%%%%%%%%%%%%%
%%%%%%%%%%%%%%%%%%%%%%%%%%%%%%%%%%%%%%%%

%Our estimates for the critical temperatures are not directly comparable to those obtained from the
%lattice simulations as in our approximation the same basis lattice is
%repeated infinitely many times, where as lattice calculations use
%periodic or free boundary conditions. However, we expect that the difference is
%uppressed with $k$. 
%If Monte Carlo calculation were approximating a
%given fractal by taking the infinite volume limit expanding instead of
%iterating the estimate of the critical temperature would agree at
%finite $k$ as pointed out

%%%%%%%%%%%%%%%%%%%%%%%%%%%%%%
\begin{center}
\begin{table}[t]
\begin{centering}
 \begin{tabular}{llll}
\hline
\hline
Authors  &  $T_C$   & $\nu$ &  k    \tabularnewline
\hline\hline
 \multicolumn{4}{c}{ \bf SC$\bf (3,1) \quad d_f = 1.8927$}\\
\hline
Bonnier \textit{et al.}\ (1987) \cite{Bonnier:1987aa}  & $1.54$ & 
$1.3$ & $3$ \\
Our work  & $1.56759$ & - & 3 \\
Pruessner \textit{et al.}\ (2001) \cite{Pruessner:2001aa}   & $1.5266(11)$ & 
$-$ & $4$ \\
Our work  & $1.525660$ & - & 4 \\
Pruessner \textit{et al.}\ (2001) \cite{Pruessner:2001aa}   & $1.5081(12)$ & 
$-$ & $5$ \\
Our work  & $1.504460$ & - & 5 \\
Pruessner \textit{et al.}\ (2001) \cite{Pruessner:2001aa}   & $1.4992(11)$ & 
$-$ & $6$ \\
Bab \textit{et al.}\ (2005) \cite{Bab:2005aa}   & $1.4945(50)$ & 
$\sim 1.39$ & $6$ \\
Our work  & $1.493310$ & - & 6 \\
Carmona \textit{et al.}\ (1998) \cite{{Carmona:1998aa}}&
$1.481$ & $1.70(1)$ &7 \\
Monceau \textit{et al.}\ (1998) \cite{Monceau:1998aa} &
$1.482(15)$ & $1.565(10)$ &7 \\
Our work  & $1.48719$ & - & 7 \\
Monceau \textit{et al.}\ (2001) \cite{Monceau:2001aa} &
$1.4795(5)$ & $>1.565$ &8 \\
\hline\hline
 \multicolumn{4}{c}{\bf SC$\bf (4,2) \quad d_f = 1.7925$}\\
\hline
Carmona \textit{et al.}\ (1998) \cite{{Carmona:1998aa}}&
$1.077$ & $3.23(8)$ &6 \\
Monceau \textit{et al.}\ (2001) \cite{Monceau:2001aa}&
$<1.049$ & $ > 3.37$ &6 \\
Our work & $1.1428$ & - & 5\\ % ix 211
\hline\hline
 \multicolumn{4}{c}{ \bf SC$\bf (5,1) \quad d_f = 1.9746$}\\
\hline
Monceau \textit{et al.}\ (2001) \cite{Monceau:2001aa}&
$2.0660(15)$ & $1.083(3)$ &5 \\
Our work  & $2.06672$ & - & 4  \\
\hline\hline
 \multicolumn{4}{c}{\bf SC $\bf (5,3) \quad d_f = 1.7227$}\\
\hline
Monceau \textit{et al.}\ (2001) \cite{Monceau:2001aa} &
$<0.808$ & $>4.06$ &5 \\
Our work & $0.974828$ & - & 4 \\
\hline\hline
Ising $2d$  & $2.269$ & $1$  & - \\
\hline 
\hline
\end{tabular}
\par\end{centering}
\caption{Comparison with the literature (Monte Carlo study).\label{tab:comparison}}
\end{table}
\end{center}
%%%%%%%%%%%%%%%%%%%%%%%%%%%%%%%%%

%%%%%%%%%%%%%%%%%%%%%%%%%%%%%%%%
\subsection{Correlation lengths}
%%%%%%%%%%%%%%%%%%%%%%%%%%%%%%%%

%
As explained in section~\ref{sub:method}, we can also  estimate the
correlation length of the system as a function of $k$, $t = T/T_C -1 $ (the
reduced temperature) and $\Lambda_k$ using Eq.~(\ref{cl}). We expect
from universality that all approximands have $\nu = 1$, since for
finite $k$, they all belong to the universality class of the two dimensional Ising model.
% $\alpha
%= 0$ (the singularity
%of the free energy is logarithmic) implies it via the hyperscaling
%relation $\alpha = 2 - d \nu$. This remains true for  any finite
%$k$.
 Different critical exponents for the fractal can emerge only in
the limit $k \to \infty$ where  the type of  singularity manifested by the free
energy at the critical point can change. 

\begin{figure}
\begin{centering}
\includegraphics[scale=1.35]{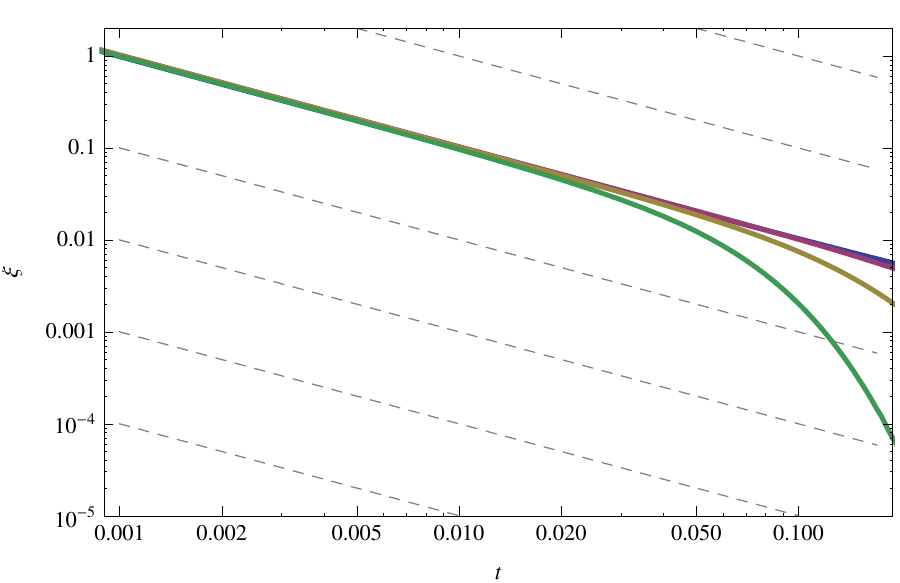}
\par
\end{centering}
\caption{Normalized correlation lengths of the Sierpinski carpet starting from
  above for $k=1,2,3,4$
   as a function of the reduced temperature $t$.
As expected from universality all all curves tend to $t^{-1}$ for small enough $t$.\label{fig:xi_vs_t}}
\end{figure}

\begin{figure}
\begin{centering}
\includegraphics[scale=1.35]{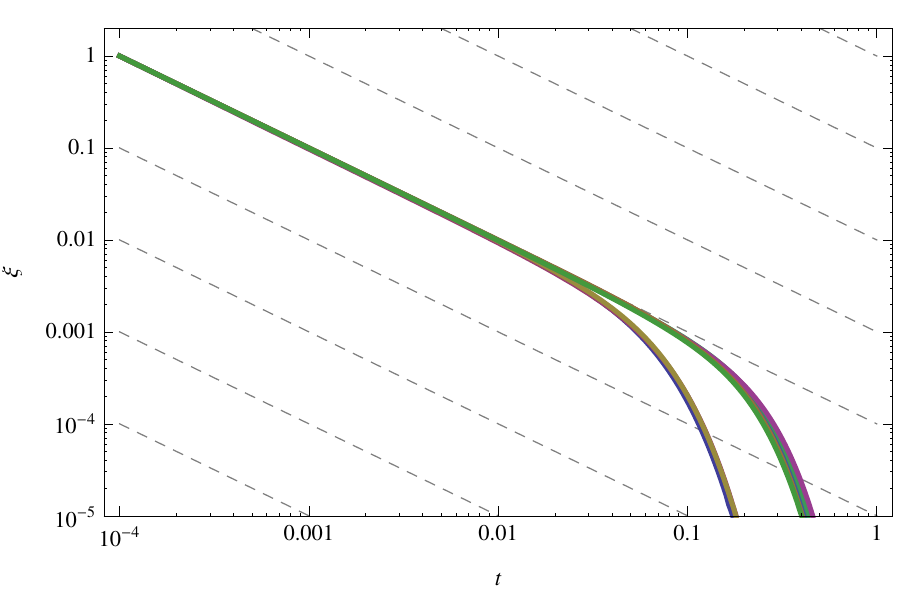}
\par
\end{centering}
\caption{Universality strikes back: normalized correlation lengths of the non-trivial $L=3$ Sierpinski carpets for $k=4$ as a function of the reduced temperature $t$.
As expected from universality all curves tend to $t^{-1}$ (dashed lines) for small enough $t$.\label{fig:xi_k4_t}}
\end{figure}

We checked numerically this expectation. Evaluating  Eq.~(\ref{cl}) exactly for various values of the reduced
temperature, we computed the correlation length for various $L=3$
fractals with $k$ up to four. 
Our results for ${\rm SC}(3,1)_k$ are illustrated in
Figure~\ref{fig:xi_vs_t} where the correlation lengths have been
normalized to exhibit universality.  As can be seen
all the curves are compatible with $\nu=1$, but the scaling region
shrinks as $k$ increases.
Our results for the normalized correlation lengths of all $L=3$ fractals considered in Table~\ref{tab:comparison}
are represented in Figure~\ref{fig:xi_k4_t}. This picture represents a strong
confirmation of universality.
%where the correlation lengths have been normalized to exhibit universality.

This analysis implies that if the limit $k\to\infty$  is continuous then
the critical exponent $\nu$ is one for the  fractals of
dimension below two. In
contrary, if the limit is discontinuous other values are possible, but
our approach cannot determine them, unless we are able to calculate
analytically the determinant $P_{\Lambda_k}(v,\bf{k})$ as we did for
the Sierpinski gaskets. Without such an analytical formula, at any
finite $k$, it is difficult to estimate the critical exponent $\nu$
of the fractals.

%%%%%%%%%%%%%%%%%%%%%
%%%%%%%%%%%%%%%%%%%%%
%
%%%%%%%%%%%%%%%%%%%%%
%%%%%%%%%%%%%%%%%%%%%
\paragraph*{Comparison with the Monte Carlo approach.}

Defining, in analogy with what done in the case of the critical temperatures, $\nu_{k,l}$ as the critical exponent for the $k^{th}$
iterations of systems of size $l$. We compute $\nu_{k,\infty}$, and
find $\nu_{k,\infty} = 1$ for all $k$ as expected from
universality. For the  fractal critical exponent $\nu_\infty
= \lim_{k \to \infty} \nu_{k,l}$ to be different from one the
limit has to be discontinuous. 
%Instead, Monte Carlo simulations compute $\nu_{k,1}$.
The standard theory of finite size scaling \cite{Fisher:1972zza} applies to changing
$l$ while keeping $k$ fixed, and this scaling should lead to $\lim_{l \to
  \infty} \nu_{k,l} = 1$ as required by universality. This was already noted by \cite{Pruessner:2001aa}.
 To our knowledge there is no scaling theory with respect to $k$ at fixed $l$, in
 particular for $l=1$, which is used in some Monte Carlo simulations. For $l=\infty$ there is not such
 theory because $\nu_{k,\infty}=1$ for all finite $k$. Hence, some Monte
 Carlo simulations rely on a possible scaling on $\nu_{k,1}$, which
 might not exist.
 
Furthermore, our results show that the scaling region where $\xi \sim
t^{-1}$ shrinks as $k$ is increased, as shown in the Figure~\ref{fig:xi_vs_t}. This suggests that it becomes increasingly
difficult to compute the critical exponent $\nu_{k,1}$ keeping $k$
fixed and calculating it using the reduced temperature as the
scaling variable, which is an other approach used in Monte Carlo
simulations. 
 
As shown in the table~\ref{tab:comparison}, Monte Carlo  simulations provide
estimates for the critical exponents, in particular $\nu$. 
They report values of $\nu>1$ even bigger than four
\cite{Monceau:2001aa}. It might be that the lattice simulations are able to capture the right
universal properties of the fractals. Nevertheless, a better
theoretical understanding of the situation is needed.

% possible issue with the lattice simulations is that they calculate the
%critical exponents using finite size scaling for systems by increasing
%$k$. However, the change of $k$ does not only alter the volume but
%also the structure of the system, which strictly speaking violates the
%scaling arguments as already noted in \cite{Pruessner:2001aa}. 
%

%%%%%%%%%%%%%%%%%%%%
\section{Conclusions and Outlook}
%%%%%%%%%%%%%%%%%%%

We showed that it is possible to approximate the solution of fractal
Ising models via a sequence of exact solutions of Ising models on
finite periodic representations of the fractal under consideration. We found that the rate of convergence to the exact solution can be very slow, as the explicit example of the exact solution of the Sierpinski gasket model shows.
We found estimates for all $L=3$ fractals, and in particular for the Sierpinsky carpet. Numerical improvements can rapidly refine our results ultimately leading to the accurate determination of the exact critical temperatures for all non-trivial fractals of dimension smaller than two.

The problem is much more difficult in the case of the critical exponents since universality ultimately sets in and renders any finite periodic approximation useless to the scope.
But we can still speculate on the actual values for the critical exponents of
the fractals. Our analysis suggests two scenarios, related to the fate
of the limits $\nu_\infty = \lim_{k \to \infty} \nu_{k,\infty}$ and
$\alpha_\infty = \lim_{k \to \infty} \alpha_{k,\infty}$. Assuming the
hyperscaling relation $\alpha =2 - d_f \nu $ to hold true for fractals with infinite ramification number, then the continuity of the first limit  implies $\nu_\infty = 1$ and a change in the singularity structure of the free energy, i.e. $\alpha_\infty = 2-d_f$. In this case both critical exponents will be continuous in the $d_f \to 1$ limit. If, instead, there is no change of singularity structure, i.e. $P_\infty(v,\bf{k})$ retains the polynomial zero it has in the finite $k$ cases, then $\alpha_\infty = 0 $ and, again assuming the hyperscaling relation, we find $\nu_\infty = 2/d_f$, which is not continuous in the $d_f \to 1$ limit but is a fairly good approximation for all {\it fractional}  dimensions as compared to the RG results \cite{Guida:1998bx,Ballhausen:2003gx}.
%; in this case both will be smooth functions in the range between one and four.
Obviously, it can also be that both limits are discontinuous or that
the hyperscaling relation is either violated or does not contain
$d_f$. It can also be that these limits are sensible to the type of
{\it fractal} under study, \textit{i.e} they depend on fine details
such as lacunarity or connectivity~\cite{gefen:1980aa,Monceau_2004}.

The question of the existence of a lower critical dimension, for fractals with infinite
ramification number (or more restrictive properties), can now  in principle be addressed by our method once the
numerical routines are improved, since we have seen that the rate of convergence
of the critical temperatures $T_k$ is slower the smaller the fractal dimension is.
High values of $k$ will thus be needed to resolve the neighbourhood of a possible lower critical dimension.

Further interesting applications of our method are related to the study of Ising models on other non-translationally
invariant lattices, like those defined on aperiodic or random lattices, or with random interactions.
  
\subsubsection*{Acknowledgments}

We thank Rudy Arthur for initial participation and discussion.
This work was supported by the Danish National Research Foundation DNRF:90 grant and by a Lundbeck Foundation Fellowship grant. The computing facilities were provided by the Danish Centre for Scientific Computing.

\end{document}